\documentclass[prb, 11pt,letterpaper,superscriptaddress,floatfix,nofootinbib,notitlepage]{revtex4-1}

\usepackage{textcomp}
\usepackage{amsmath}
\usepackage{amsfonts}
\usepackage{amssymb}
\usepackage{graphicx}
\usepackage{siunitx}
\usepackage{caption}
\usepackage{setspace}
\usepackage{xcolor}
\setcitestyle{super}
\usepackage[colorlinks, citecolor={blue}, linkcolor={black}]{hyperref}

\begin{document}
\raggedbottom

\author{Jeong Min Park}
\thanks{These authors contributed equally}
\affiliation{Department of Physics, Massachusetts Institute of Technology, Cambridge, Massachusetts 02139, USA}

\author{$^{\!\!\!\!,\ \!\dagger}$ Yuan Cao}
\thanks{These authors contributed equally}
\affiliation{Department of Physics, Massachusetts Institute of Technology, Cambridge, Massachusetts 02139, USA}
\affiliation{Department of Physics, Harvard University, Cambridge, Massachusetts 02138, USA}

\author{Liqiao Xia}
\affiliation{Department of Physics, Massachusetts Institute of Technology, Cambridge, Massachusetts 02139, USA}

\author{Shuwen Sun}
\affiliation{Department of Physics, Massachusetts Institute of Technology, Cambridge, Massachusetts 02139, USA}

\author{Kenji Watanabe}
\author{Takashi Taniguchi}
\affiliation{National Institute for Materials Science, Namiki 1-1, Tsukuba, Ibaraki 305-0044, Japan}
 
\author{Pablo Jarillo-Herrero}
\email{parkjane@mit.edu, pjarillo@mit.edu}
\affiliation{Department of Physics, Massachusetts Institute of Technology, Cambridge, Massachusetts 02139, USA}
 
\title{Magic-Angle Multilayer Graphene: A Robust Family of Moiré Superconductors}

\maketitle

\textbf{The discovery of correlated states and superconductivity in magic-angle twisted bilayer graphene (MATBG)\cite{cao_correlated_2018, cao_unconventional_2018} has established moir\'e quantum matter as a new platform to explore interaction-driven and topological quantum phenomena\cite{andrei_marvels_2021}. Multitudes of phases have been realized in moir\'e systems, but surprisingly, robust superconductivity has been one of the least common of all, initially found in MATBG and only more recently also in magic-angle twisted trilayer graphene (MATTG)\cite{park_tunable_2021, hao_electric_2021}. While MATBG and MATTG share some similar characteristics, they also exhibit substantial differences, such as in their response to external electric and magnetic fields. This raises the question of whether they are simply two separate unique systems, or whether they form part of a broader family of superconducting materials. Here, we report the experimental realization of magic-angle twisted 4-layer and 5-layer graphene (MAT4G and MAT5G, respectively), which turn out to be superconductors, hence establishing alternating-twist magic-angle multilayer graphene\cite{khalaf_magic_2019} as a robust family of moir\'e superconductors. 
The members of this family have flat bands in their electronic structure as a common feature, suggesting their central role in the observed robust superconductivity. On the other hand, there are also important variations across the family, such as different symmetries for members with even and odd number of layers.
However, our measurements in parallel magnetic fields, in particular the investigation of Pauli limit violation and spontaneous rotational symmetry breaking, reveal that the most pronounced distinction is between the $N=2$ and $N>2$-layer structures. Our results expand the emergent family of moiré superconductors, providing new insight with potential implications for the design of novel superconducting materials platforms.}

Moir\'e quantum matter results from stacking two or more atomically thin materials with a lattice mismatch or at a relative twist angle\cite{andrei_marvels_2021}. Since the discovery of MATBG\cite{cao_correlated_2018,cao_unconventional_2018}, moir\'e systems have actively expanded to include different types of constituent layers and structures. Correlated and topological phenomena including but not limited to correlated insulators, quantum anomalous Hall effects, ferromagnetism, and generalized Wigner crystals have been discovered and reproduced in various new moiré systems thus far\cite{chen_evidence_2019, burg_correlated_2019, shen_correlated_2020, cao_tunable_2020, liu_tunable_2020, he_symmetry_2020, polshyn_electrical_2020, xu_tunable_2021, chen_electrically_2020, regan_mott_2020, tang_simulation_2020, wang_correlated_2020, xu_correlated_2020, jin_stripe_2021,li_continuous_2021,li_quantum_2021}. However, robust and reproducible moiré superconductivity was only seen in MATBG for the first few years\cite{cao_unconventional_2018,yankowitz_tuning_2019, lu_superconductors_2019}, despite reports of signatures of superconductivity in a few other systems \cite{chen_signatures_2019, burg_correlated_2019, shen_correlated_2020, liu_tunable_2020,he_symmetry_2020,wang_correlated_2020, xu_tunable_2021, zhang_correlated_2021}.

More recently, robust and highly tunable superconductivity has been demonstrated in MATTG\cite{park_tunable_2021, hao_electric_2021}. Remarkably, the superconductivity in MATTG persists up to in-plane magnetic fields $\sim3$ times larger than the Pauli limit for conventional BCS superconductors\cite{cao_pauli-limit_2021}, whereas the critical magnetic field in MATBG did not significantly violate its nominal Pauli limit\cite{cao_unconventional_2018}. The similarities and differences between MATBG and MATTG raise the question of what the key ingredients needed to realize robust moiré superconductivity are, and whether the two systems may be part of a larger family of novel superconductors. Practically, it is desirable to find a reliable way to construct new moir\'e superconductors, since the existence of flat bands alone does not guarantee superconductivity, as demonstrated in several other graphene-based moir\'e systems\cite{cao_tunable_2020, liu_tunable_2020, he_symmetry_2020,polshyn_electrical_2020,xu_tunable_2021,chen_electrically_2020}. Such quest could also substantially help understand the mechanism underlying these superconductors. 

\begin{figure}
\includegraphics[width=1\textwidth]{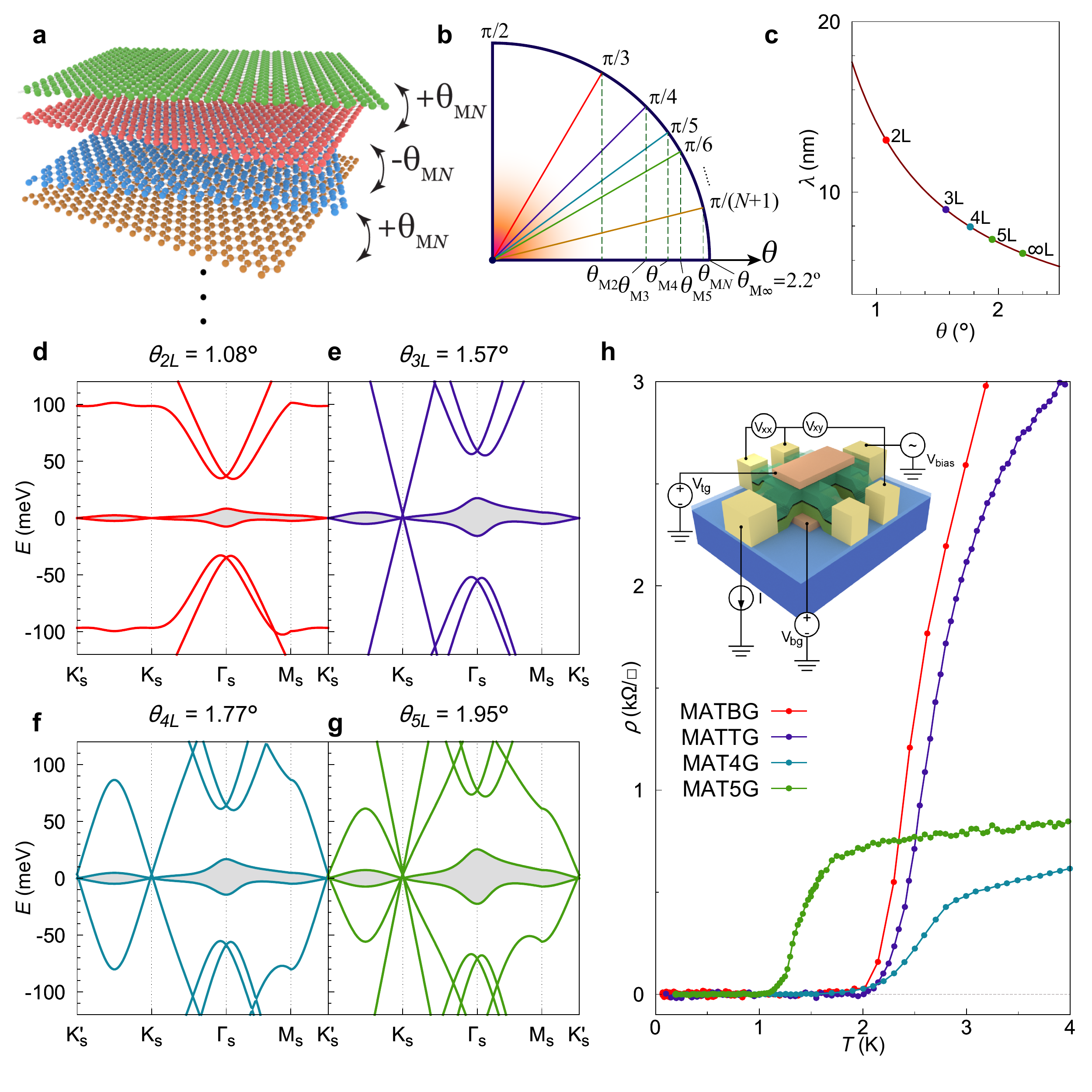}
\caption{Magic-angle multilayer graphene. (a) Twisted multilayer graphene with alternating twist angles $\theta_{\mathrm{M}N}$ and $-\theta_{\mathrm{M}N}$ between the adjacent layers, where $\theta_{\mathrm{M}N}$ is the magic angle $\theta_{\mathrm{M}}$ specific to an $N$-layer structure. (b) In the chiral limit, $\theta_{\mathrm{M}N}$ can be obtained for any $N$ from the asymptotic value $\theta_{\mathrm{M}\infty}=\SI{2.2}{\degree}$, by a simple trigonometric transformation. (c) Dependence of the moir\'e wavelength $\lambda$ on the twist angle. Note that we only consider structures with atomic alignment between the $n$-th and $(n+2)$-th layers, so that a single moir\'e wavelength can be defined\cite{khalaf_magic_2019}. (d-g) Single-particle band structures for TBG, TTG, T4G, and T5G, respectively, at representative angles near their respective magic angle. The flat bands that are shared by all systems are color-coded with gray. The flat bands in MATBG are isolated from all other dispersive bands by band insulators, whereas $N>2$ structures have extra bands coexisting with them. The extra bands consist of either pristine Dirac-like bands or non-magic-angle-like TBG bands, depending on $N$. (h) Resistivity $\rho$ versus temperature $T$ curves for MATBG, MATTG, MAT4G and MAT5G ($N=2,3,4,5$), showing superconducting transitions in all four systems at their respective magic angle. Their twist angles correspond to the same values used for the calculations in Fig. 1d-g, which are indicated in Fig. 1c as well. The normal-state resistivity of MAT4G and MAT5G are smaller than MATBG and MATTG, likely due to the presence of the extra dispersive bands.} 
\end{figure}

\subsection{Magic-angle multilayer graphene}

When two layers of monolayer graphene (MLG) are twisted at a small angle\cite{suarez_morell_flat_2010,bistritzer_moire_2011,lopes_dos_santos_continuum_2012}, hybridization between the Dirac bands in the graphene layers can give rise to unique flat bands where the Fermi velocity vanishes. This happens when the twist angle is close to a series of `magic' angles. Such twisted bilayer graphene structure, with the first `magic' angle of around \SI{1.1}{\degree}, has been intensively studied, providing insights into the nature of the correlated states, nontrivial topology, and superconductivity\cite{cao_correlated_2018,cao_unconventional_2018, yankowitz_tuning_2019, lu_superconductors_2019, xie_spectroscopic_2019, wong_cascade_2020, zondiner_cascade_2020, park_flavour_2021, oh_evidence_2021, pierce_unconventional_2021, xie_fractional_2021}. It has been theoretically shown\cite{khalaf_magic_2019} that for three or more twisted layers of graphene, there are similar series of `magic' angles if the layers are alternatively twisted by $(\theta, -\theta, \theta, \ldots)$ (Fig. 1a). The values of such angles can be analytically computed from the bilayer value in the chiral limit, where the interlayer hopping at AA sites is turned off\cite{khalaf_magic_2019}. As illustrated in Fig. 1b, they are in fact elegantly related by simple trigonometric transformations, i.e. the largest magic angle can be expressed as $\mathrm{\theta_{N}=\theta_{\infty}\cos\frac{\pi}{N+1}}$, where $N$ is the number of layers, and $\theta_{\infty}=2\theta_{N=2}$ is the asymptotic limit of the largest magic angle as $N\rightarrow\infty$. As $N$ increases, the magic angle increases and the moiré length scale decreases (Fig. 1c). The real magic angle values slightly deviate from the values in the chiral limit. For 3, 4, and 5 layers (3L, 4L, and 5L), the largest magic angles are located at approximately \SI{1.57}{\degree}, \SI{1.77}{\degree} and \SI{1.95}{\degree}, respectively. Figure 1d-g shows the electronic bands corresponding to 2L$\sim$5L calculated using a continuum model\cite{bistritzer_moire_2011}. Notably, all these `magic' structures host a pair of flat bands with extremely small dispersion. MATBG, which is the first one in the series, has a single pair of flat bands near zero energy that are isolated from all other dispersive bands, whereas for the structures with $N>2$ layers, there are extra bands that form $N-2$ additional Dirac points at low energies (per valley and spin). Due to the presence of these extra bands in $N>2$ layers, the electronic structures are strongly modified upon application of an out-of-plane displacement field, which tends to hybridize the flat bands with other dispersive bands (Extended Data Fig. 1). 

This `family' of magic-angle moir\'e structures shares a number of common properties. Regardless of the number of layers, these structures have a single moir\'e periodicity determined by $\theta$, and each flat band hosts a density of $n_s=\sqrt{3}\theta^2/8a^2$ (including valley and spin degeneracies), where $a=\SI{0.246}{\nano\meter}$ is the lattice constant of graphene. It is therefore convenient to use $\nu=4n/n_s, -4<\nu<4$ to describe the carrier density $n$ within the flat bands. All members of the family globally retain the $C_{2z}$ symmetry of graphene (see Methods for discussion about atomic alignment). For odd $N$, the atomic structure has an out-of-plane mirror symmetry $M_z$, whereas for even $N$ this is replaced by a $C_{2}$ rotation axis that lies in the $x$-$y$ plane. In previous works\cite{cao_correlated_2018, cao_unconventional_2018, park_tunable_2021, hao_electric_2021}, MATBG and MATTG have both been shown to exhibit correlated insulator/resistive states at $\nu=\pm2$, as well as superconductivity in the vicinity of these states, with critical temperatures up to $\sim$\SI{3}{\kelvin}. Since the flat bands in magic-angle structures with $N>2$ can be mathematically mapped onto those in $N=2$ (MATBG), one may expect them to be potential host of superconductivity as well.

\begin{figure}
\includegraphics[width=1\textwidth]{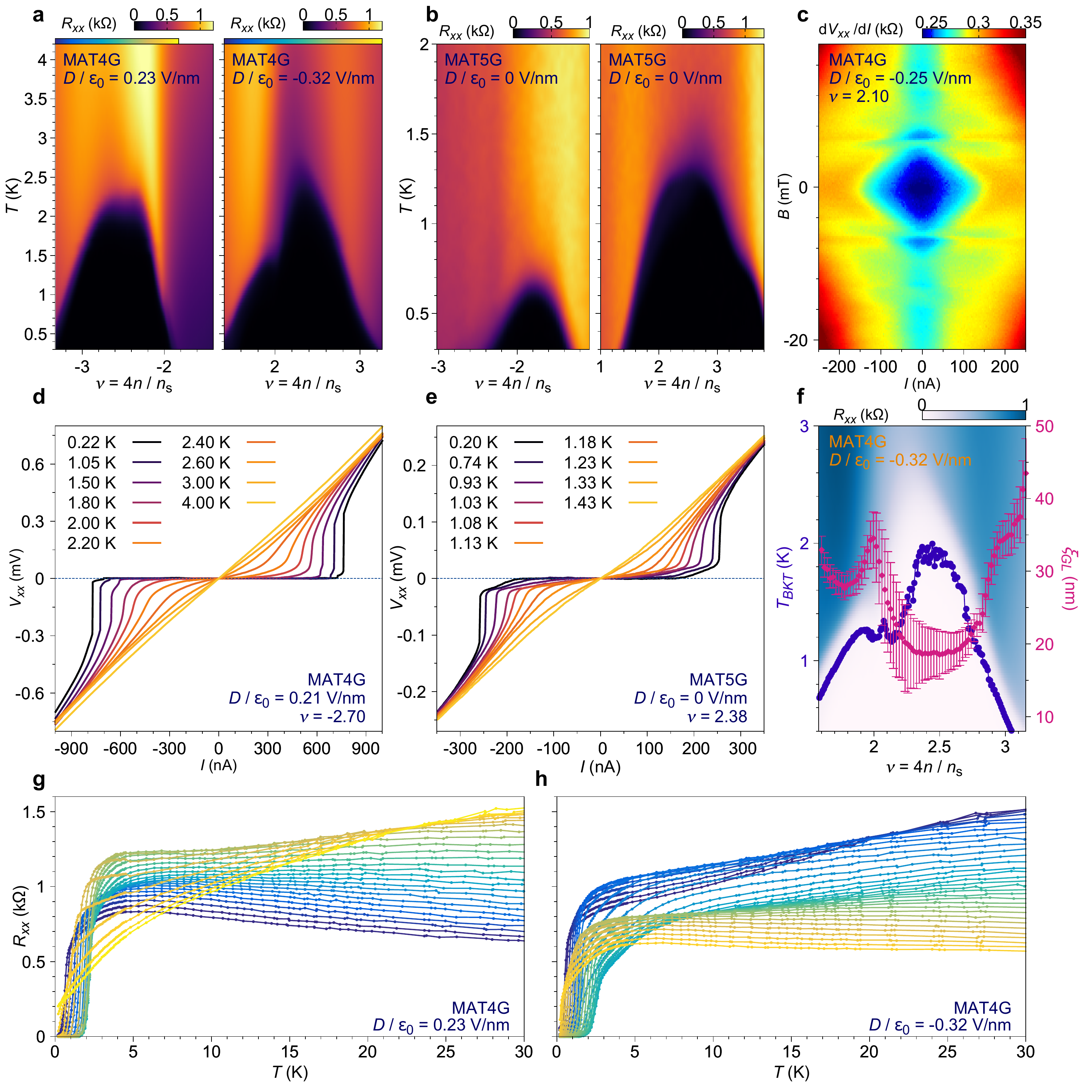}
\caption{\small Robust superconductivity in MAT4G and MAT5G. (a-b) Resistance $R_{xx}$ versus moir\'e filling factor $\nu$ and temperature $T$ for MAT4G and MAT5G, respectively. The superconducting domes span a wide density range across the flat bands. Note that in MATTG, MAT4G, and MAT5G, $\nu$ includes the filling of both the flat bands and the extra dispersive bands. (c) Differential resistance $dV_{xx}/dI$ versus dc bias current $I$ and small perpendicular magnetic field $B_{\perp}$, showing Fraunhofer-like oscillations in $B_{\perp}$. The data are measured in a split top gate geometry where the middle non-gated region is tuned to a resistive state (N) while maintaining the gated regions superconducting (S), thus forming an SNS Josephson junction\cite{rodan-legrain_highly_2021, de_vries_gate-defined_2021}. (d) Voltage ($V$) versus current ($I$) curves at $T$ ranging from \SI{220}{\milli\kelvin} to \SI{4}{\kelvin} at $\nu=-2.70$ and $D/\varepsilon_0=\SI{0.21}{\volt\per\nano\meter}$ in MAT4G. We find sharp transitions at low $T$, with a weakly linear in $T$ dependence above $\sim \SI{4}{\kelvin}$. (e) Same measurement in MAT5G at $\nu=2.38$ and $D/\varepsilon_0=\SI{0}{\volt\per\nano\meter}$. (f) Ginzburg-Landau coherence length $\xi_{GL}$ versus $\nu$ at $D/\varepsilon_0=\SI{-0.32}{\volt\per\nano\meter}$ in MAT4G, along with the extracted Berezinskii–Kosterlitz–Thouless transition temperature $T_{BKT}$. They are overlaid on a colormap of $R_{xx}$ versus $\nu$ and $T$. $\xi_{GL}$ reaches low values around \SI{20}{\nano\meter} near optimal doping $\nu\sim 2.5$. The extraction is performed with $25\%$, $30\%$, and $35\%$ of normal-state resistance for the lower uncertainty bound, middle value, and upper uncertainty bound, respectively (see Methods). (g-h) $R_{xx}$ versus $T$ curves in MAT4G at $D/\varepsilon_0=\SI{0.23}{\volt\per\nano\meter}$ (g) and $D/\varepsilon_0=\SI{-0.32}{\volt\per\nano\meter}$ (h) across $\nu$ up to $T=\SI{30}{\kelvin}$ showing sharp superconducting transitions. The color scale for the curves matches the scale bar shown in the hole-doped and electron-doped plots at the top in (a), respectively.} 
\end{figure}

\subsection{A family of robust superconductors}

We fabricated and measured high quality magic-angle tetralayer and pentalayer graphene devices (MAT4G and MAT5G, respectively) and observed robust superconductivity in both systems, thus establishing alternating twist magic angle multilayer graphene as a new `family' of robust moir\'e superconductors. We note that we have fabricated and measured three MAT4G and four MAT5G devices and, remarkably, all of them showed robust superconductivity, hence yielding a very high device success rate (see Methods for details). Our device yield for MATTG superconducting devices is also very high, which may indicate that the superconducting phase in magic-angle systems with $N>2$ may be less susceptible to relaxation or disorder (e.g. twist angle disorder) than it is the case for MATBG (for which our yield is considerably lower, about 50\%). Figure 1h shows representative resistivity versus temperature curves for all four members of the family. These curves have been chosen for filling factors and displacement fields where $T_{c,50\%}$ was maximum for each device. In particular, for the new members MAT4G and MAT5G, $T_{c,50\%}$ is $\sim$\SI{2.76}{\kelvin} and $\sim$\SI{1.38}{\kelvin}, respectively.  

The normal-state resistivity in MAT4G and MAT5G is considerably lower than that in MATBG and MATTG, which is possibly due to the presence of extra highly dispersive Dirac bands, which provide parallel conducting channels. Figure 2a-b shows the $\nu$-$T$ phase diagrams of MAT4G and MAT5G, respectively (see Extended Data Fig. 3 for other devices). The range of filling factors in which the superconductivity appears in MAT4G and MAT5G is generally wider than in MATTG and MATBG, starting close to $\nu=\pm1$ and reaching beyond $\nu=\pm3$. In particular, superconductivity in MAT5G extends to or can even reach beyond $\nu=+4$ (see Extended Data Fig. 1 for the other device). Considering that MATTG also had a wider dome compared to MATBG\cite{park_tunable_2021,hao_electric_2021}, this observation suggests that increasing the number of layers could possibly increasing the phase space robustness of the superconductivity. However, one should also note that for $N>2$, $\nu$ does not indicate the filling factor of the flat bands, because some of the carriers induced by the gates fill the dispersive bands. This effect should be more pronounced as $N$ increases, since the number of additional dispersive Dirac bands is $N-2$. Moreover, as $N$ increases, an inhomogeneous distribution of charge carriers among the layers could alter the effective filling factor in the flat band (see Methods and Extended Data Fig.4). In addition, the correlated resistive states at $\nu=\pm2$, if present, are less resistive than those  in MATTG\cite{park_tunable_2021}, in some cases even absent in the phase diagram (see Extended Data Fig. 1 for the full $\nu-D$ phase space for the systems), in contrast to the relatively highly insulating states observed in MATBG\cite{cao_correlated_2018, yankowitz_tuning_2019, lu_superconductors_2019}. This trend again might be attributed to the presence of additional Dirac bands at the Fermi level corresponding to $\nu=\pm2$ as $N$ increases. The presence of such bands would make the overall structure gapless even if the flat band opens a correlated gap. 

To further confirm the superconductivity in the `magic' family, we have measured the voltage-current ($V-I$) characteristics in both MAT4G and MAT5G (Fig. 2d-e). The sharp switching behaviour in the $V-I$ curves in all the devices (see Extended Data Fig. 3) confirms the true, robust superconductivity in these new members of the family. In addition, we have also performed measurements of the critical current versus perpendicular magnetic field, which reveal a Fraunhofer-like oscillation pattern (Fig. 2c, for MAT4G device).  We note that due to the absence of strongly insulating states in these multilayer systems, the Fraunhofer-like pattern could only be obtained by constructing a gate-defined Josephson junction, as previously done in MATBG\cite{rodan-legrain_highly_2021, de_vries_gate-defined_2021}. The Ginzburg-Landau coherence length measured in MAT4G (see Extended Data Fig. 3 for other devices) is short and around \SI{20}{\nano\meter}, suggesting a relatively strong coupling, as observed in MATTG \cite{park_tunable_2021}. Similar to MATTG, all of these properties are further tunable upon application of electric displacement field $D$ (see Extended Data Fig. 1 for the full $\nu$-$D$ map). Extending the measurements to higher temperature (Fig. 2g-h), we find that the superconducting transition in these systems is relatively sharp, especially compared to MATBG, which typically exhibits very broad transitions. At higher temperatures, the resistivity does not show a strong temperature dependence, unlike the steep linear-in-temperature behavior previously found in MATBG\cite{cao_strange_2020, polshyn_large_2019}. The weak linear-in-T behaviour observed might be the result of contributions stemming from both the flat bands and dispersive bands (the latter being very weakly T dependent \cite{dean_boron_2010}), although further theoretical work and experiments are needed to determine if there are signatures of strange metal behavior in these large $N$ devices.

\begin{figure}
\includegraphics[width=1\textwidth]{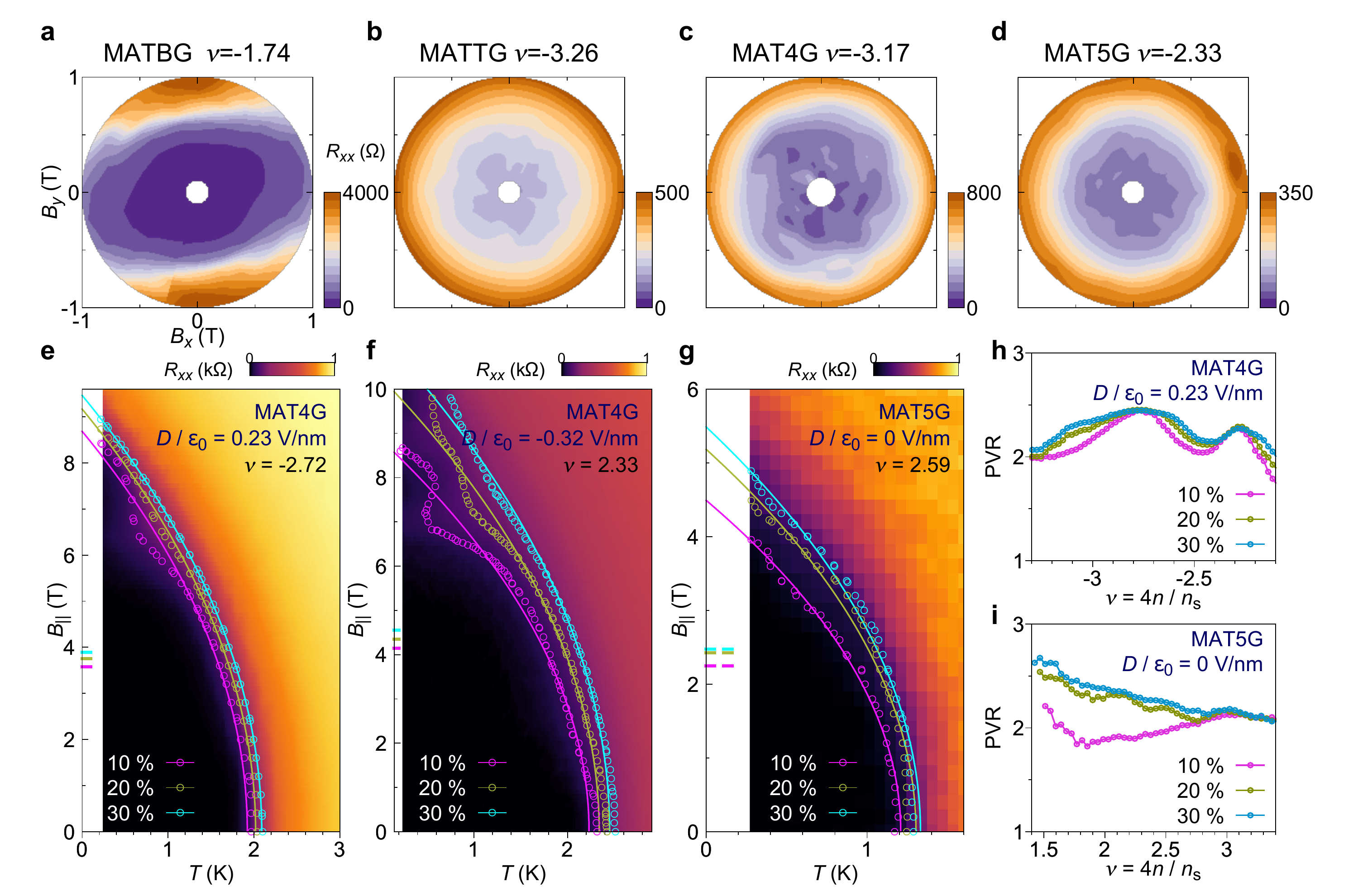}
\caption{In-plane magnetic field dependence of the superconducting states. (a-d) Polar maps of the in-plane magnetic field response of $R_{xx}$ in MATBG, MATTG, MAT4G ($\theta=\SI{1.7}{\degree}$; different device from Fig. 1,2,and 3e-f), and MAT5G, at $\nu=-1.74,-3.26,-3.17, -2.33$ and $T=\SI{0.07}{\kelvin}, \SI{0.1}{\kelvin}, \SI{0.2}{\kelvin}, \SI{0.2}{\kelvin}$, respectively. MATBG shows an anisotropic response with two-fold symmetry in its in-plane critical field, while the other three systems do not show any anisotropy. (e-g) $B_\parallel$-$T$ phase diagram at $(\nu,D/\varepsilon_0)=$ $(-2.72,\SI{0.23}{\volt\per\nano\meter})$ (e) and $(2.33,\SI{-0.32}{\volt\per\nano\meter})$ (f) in MAT4G, and $(2.59,\SI{0}{\volt\per\nano\meter})$ in MAT5G (g). The data points denote constant-resistance contours at \SI{10}{\percent}, \SI{20}{\percent}, and \SI{30}{\percent} of the zero-field normal state resistance. Solid curves are fit to the Ginzburg-Landau expression $T\propto 1-\alpha B_\parallel^2$ ($\alpha$ is a fitting parameter). We find the critical magnetic fields $B_{c\parallel,0}$ by extrapolating the contours to zero temperature. The colored ticks on the $B_{\parallel}$-axis represent the corresponding nominal Pauli limit for each threshold. We note that in (e-f) some hints of re-entrant superconducting behaviour at high field can be appreciated\cite{cao_pauli-limit_2021}. (h-i) Pauli violation ratio (PVR), the ratio between $B_{c\parallel,0}$ and the nominal Pauli limit, across $\nu$ in MAT4G and MAT5G, respectively. In both systems, PVR is around $2\sim 3$.} 
\end{figure}

\subsection{Magnetic Field and Orbital Effects} 

One way to obtain deeper insights into the underlying mechanisms and possible differences between the family members is through the response to magnetic fields applied parallel to the two-dimensional (2D) plane of the sample ($B_{\parallel}$). Typically, magnetic fields suppress superconductivity either by inducing vortices or by closing the gap via the Zeeman effect acting on the spin component of the Cooper pairs. The former effect is largely absent for $B_{\parallel}$ applied to a 2D superconductor, whereas the latter effect leads to a nominal Pauli paramagnetic limit, $B_P=(\SI{1.86}{\tesla\per\kelvin})\times T_c$ for conventional spin-singlet superconductors with negligible spin-orbit interactions. In MATBG, it has been shown that the critical in-plane magnetic field $B_{c\parallel}$ is not substantially larger than $B_P$, and the superconductivity vanishes around such field\cite{cao_unconventional_2018}. In MATTG, on the other hand, the effect of  $B_{\parallel}$ is much weaker\cite{cao_pauli-limit_2021}, and superconductivity can persist up to fields $\sim3$ times larger than the nominal Pauli limit. This large discrepancy between MATBG and MATTG, which are close siblings in the family, may originate for a variety of reasons, including the role of in-plane orbital effects\cite{qin_-plane_2021}, a difference in superconducting pairing symmetry, and/or different Cooper pair spin configurations. Moreover, the response of the superconducting state in MATBG under different $B_{\parallel}$ directions shows an interesting spontaneous breaking of rotational symmetry\cite{cao_nematicity_2021}. While the moir\'e lattice in MATBG possesses a sixfold rotational symmetry, $B_{c\parallel}$ shows only a two-fold symmetry (Fig. 3a), suggesting that a spontaneous nematic ordering occurs in the superconducting state. Examining these phenomena across other members of the family could thus help elucidate their underlying origin and provide information regarding the nature of the superconductivity. 

Figures 3b-d shows longitudinal resistance $R_{xx}$ as a function of the magnitude and direction of $B_{\parallel}$ up to $\SI{1}{\tesla}$, for MATTG, MAT4G and MAT5G, respectively (see Methods for sample tilt calibration details). In all three systems, the superconductor to normal state transition does not display any visible dependence on the direction of $B_{\parallel}$, which is evident from the fact that the contours at different resistance values are all roughly circular (with random irregularities due to measurement noise). This is in stark contrast with MATBG (Fig. 3a), where elongated elliptical contours can be clearly seen, indicating a two-fold anisotropy of the $B_{c\parallel}$. We note that these measurements are taken near the boundary of the superconducting domes, since at optimal doping the superconducting state does not turn normal even when $B_\parallel=\SI{1}{\tesla}$ is applied, which is the highest available field in our vector magnet for the sample mounting configuration required for measuring the angle-dependent critical field.

In order to obtain $B_{c\parallel}$ at base temperature, we rotated the samples so that an in-plane field up to \SI{10}{\tesla} could be applied. These high-field measurements reveal violation of the Pauli limit in both MAT4G and MAT5G, to a similar extent as in MATTG\cite{cao_pauli-limit_2021}. Figure 3e-g shows $R_{xx}$ versus $B_{\parallel}$ and $T$ for the hole-doped and electron-doped sides of MAT4G and electron-doped side of MAT5G, respectively, with the constant-$R_{xx}$ contour and their respective fit to the Ginzburg-Landau expression $T\propto 1-\alpha B_{\parallel}^2$, where $\alpha$ is a fitting parameter  (see Methods for details of fitting, and Extended Data Fig. 5 for additional data). Three different contours at $10\%$, $20\%$ and $30\%$ of the normal-state $R_{xx}$ were chosen for the analysis. The zero-temperature critical field $B_{c\parallel}(0)$ obtained by extrapolating the fit shows values that are consistently exceeding the Pauli limit by a factor of $\sim2$ in all samples that we measured, as well as across the entire superconducting domes, as shown in Fig. 3h-i. Such consistency suggests that the Pauli limit violation is likely inherent to the superconducting state in MATTG, MAT4G, and MAT5G, instead of being due to spin-orbit coupling or strong coupling effects (see ref:\cite{cao_pauli-limit_2021} for a discussion of these effects).

\begin{figure}
\includegraphics[width=1\textwidth]{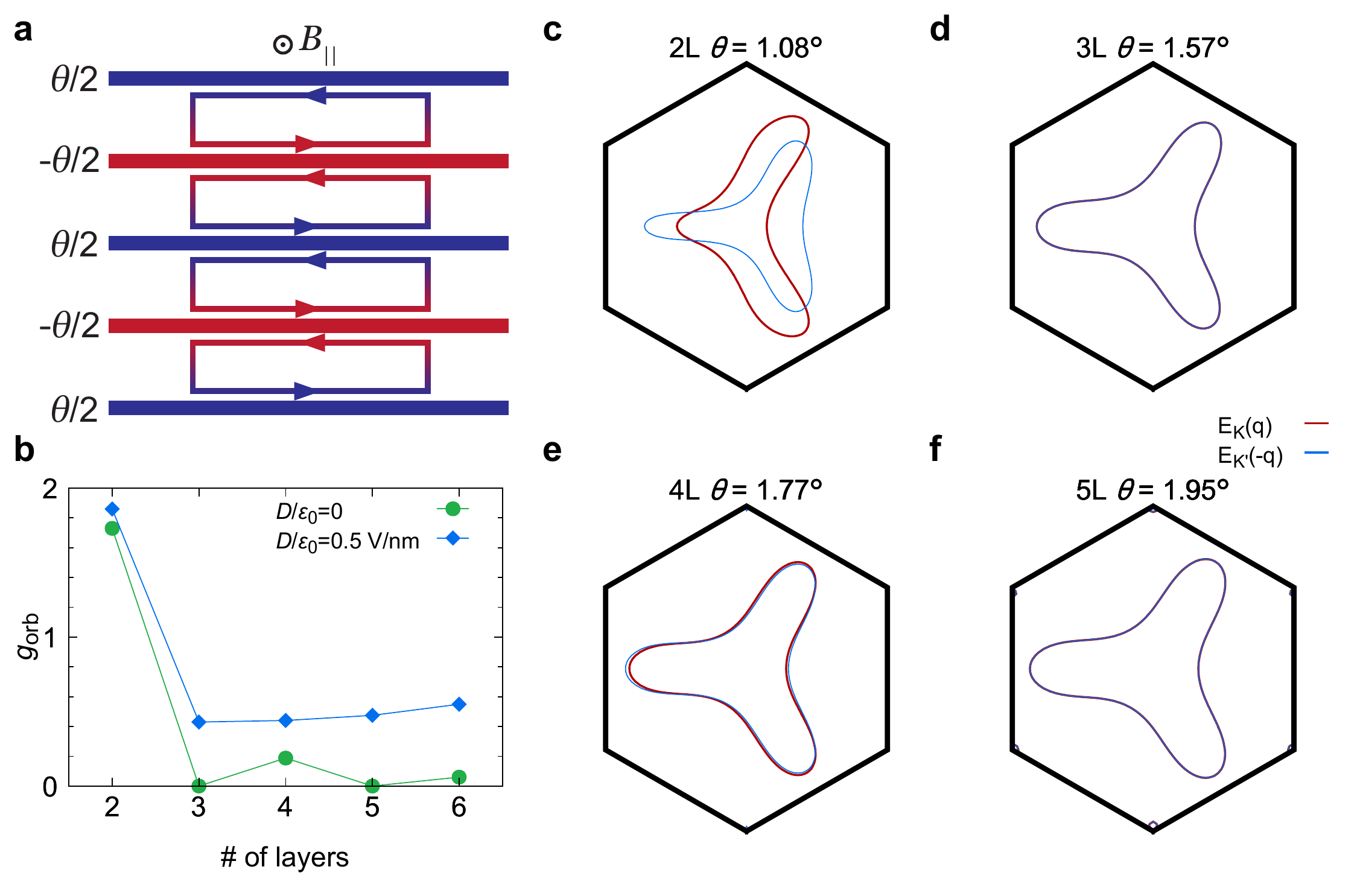}
\caption{In-plane magnetic field orbital effect. (a) Schematic showing the cancellation of the orbital effect under $B_{\parallel}$ in the alternating-twist structures. Solid blue and red lines represent graphene layers with alternating twist angles, and the arrowed-loops show that the direction of the momentum boost for hopping between adjacent pairs of layers is opposite. For the internal layers in $N>2$ structures, this results in a greatly reduced in-plane orbital effect (see Methods for mathematical derivation). (b) Calculated orbital $g$-factor, $g_\mathrm{orb}$, for $N$ layer MATNG. Both at zero and finite $D$, MATBG has the highest coupling to the in-plane field, whereas systems with $N>2$ have much smaller $g_\mathrm{orb}$. Finite $D$ breaks $M_z$ and leads to increased $g_{orb}$ compared to the case of $D=0$. (c-f) Calculated Fermi contours at $K$ and $K'$ valleys of MATBG, MATTG, MAT4G, and MAT5G, respectively, near their magic angles under $B_{\parallel}$. The magnetic field is along the horizontal direction, and the magnitude is set to \SI{20}{\tesla} to exaggerate the effect. The Fermi surface is distorted differently for K and K' valleys. For MATBG, this leads to a considerable orbital pair-breaking effect, whereas for $N>2$ structures such distortion is minimal.} 
\end{figure}

Our experiments clearly establish that, while all members of the magic-angle graphene family show similarities which likely originate from their respective flat band physics, such as the range of density where superconductivity is strongest, the in-plane magnetic field response sharply distinguishes $N=2$ (i.e. MATBG) from family members with $N > 2$. This is surprising, as from a symmetry point of view, members with even $N$ share the same in-plane $C_2$ rotation symmetry and members with odd $N$ share the same mirror symmetry $M_z$. One would thus expect that systems with even layers and odd layers behave similarly within their respective groups, while distinct across them. These observations do not depend substantially on the presence or absence of a displacement field (see Methods and Extended Data Fig. 6), and are therefore necessarily intrinsic to their respective flat bands, even though all of them can be mapped to the MATBG ones\cite{khalaf_magic_2019}.

Here we attempt to give a possible unified explanation of these experimental findings by considering orbital effects in an in-plane magnetic field. Both the strong violation of the nominal Pauli limit and the absence of nematicity in the magic-angle structures with $N>2$ layers could be accounted for by a reduced in-plane orbital effect as a result of symmetry. In 2D superconductors like the magic-angle family, while $B_\parallel$ does not induce vortices, the effective momentum boost experienced when electrons tunnel between different layers (proportional to $B_\parallel$ and the interlayer distance) can distort the shape of the Fermi surface, and this can act as a pair-breaking effect for superconductivity. This was considered for MATBG \cite{cao_nematicity_2021, qin_-plane_2021} and the pair-breaking effect has a similar magnitude as that resulting from spin Zeeman effect, with an effective $g$-factor around 2. In other words, the in-plane orbital effect in MATBG could account for the suppression of superconductivity at fields of the order of the nominal Pauli limit.

When more layers are added, however, the in-plane orbital effect between layers tends to cancel each other, instead of being added up. This is conceptually sketched in Fig. 4a. If we consider any internal layer in the stack, e.g. a layer twisted at $-\theta/2$, the electrons that tunnel from the layer above it and below it (both of which are at angle $\theta/2$) would experience opposite momentum boost. From a perturbation theory point of view, the first-order effect on the spectrum, proportional to $B_\parallel$, would partially or fully cancel depending on $N$ (see Methods for full derivation). In Fig. 4b, we calculated the mean orbital $g$-factor, $g_{orb}$, in the flat bands of magic-angle structures with $N=2,3,4,5,6$, in the absence and presence of a displacement field (see Methods; note that a similar calculation and interpretation was recently made independently in a theoretical work\cite{ledwith_tb_2021}). To interpret the role of $g_{orb}$, a system with $g_{orb}=2$ would have its critical field at the Pauli limit, and $g_{orb}<2$ leads to Pauli limit violation. MATBG has a $g_{orb}\lesssim 2$. For odd $N$, the mirror symmetry ($M_z$) of the system prohibits the in-plane magnetic fields from coupling to the system\cite{qin_-plane_2021, lake_reentrant_2021}, and $g_{orb}$ is strictly zero in the absence of displacement field. For even $N\geq 4$, although a finite value of $g_{orb}$ is allowed by symmetry, our calculations show that $g_{orb}$ is greatly suppressed compared to MATBG. Therefore, the trend of $g_{orb}$ with $N$ qualitatively explains why all $N>2$ structures violate the Pauli limit while $N=2$ does not. When a displacement field is turned on, we find that the difference between even/odd $N$ diminishes, but the orbital $g$-factors for all $N>2$ structures are still considerably smaller than that in MATBG. We note that as $N$ increases, the distribution of the displacement field becomes highly nonuniform across the stack due to electrostatic screening from outer layers, and we have taken a simple approximation to account for such effect (see Methods).

Such difference in the orbital coupling might also determine the appearance of nematicity throughout the family. Figure 4c-f shows the calculated Fermi surfaces of $K$ and $K'$ valleys upon application of $B_{\parallel}$. The momentum boosts induced by $B_{\parallel}$ are opposite for K and K' valleys. For $N=2$ (MATBG), this results in a sizeable distortion of the Fermi surface, which provides a means for $B_\parallel$ to couple to the superconducting order parameter, and this coupling is dependent on the direction of $B_\parallel$. This could lead to the observed two-fold nematicity when further pinning from strain or other many-body effects like density waves are present\cite{cao_nematicity_2021}. In contrast, for the $N>2$ structures, the Fermi surfaces in the two valleys have minimal distortion (without displacement field), and thus there is no direct coupling between $B_\parallel$ and the orbital part of the superconducting order parameter. While $B_\parallel$ could still couple to the spin degree of freedom, weak spin-orbit interaction in these graphene systems prevents coupling to the orbital part, and hence no nematicity or other types of anisotropic phases are likely to develop.

In the above discussions, the role of electron spins were deliberately neglected, which we can consider now. For $N>3$, the Cooper pairs clearly cannot be spin-singlets, or the Zeeman effect would suppress superconductivity at the Pauli limit. Since the superconducting states are otherwise similar across the family, there is a possibility that even MATBG is a non-spin-singlet superconductor, and the apparent conformance to the Pauli limit could be a result of the orbital effect as discussed above, with  similar $g_{orb}$ as for the spin Zeeman effect. Similar considerations were proposed in a recent theoretical work\cite{qin_-plane_2021}.

We note that although this mechanism can qualitatively capture the effect of in-plane magnetic fields on the magic-angle superconductor family, there are still remaining puzzles. While the $g_{orb}$ for odd $N$ are expected to be zero and the critical magnetic field should in principle be infinitely large, experimentally we find Pauli limit violation of $\sim3$ in MATTG and $\sim2$ in MAT5G. Furthermore, the theoretically calculated $g_{orb}$ strongly depends on the applied displacement field, whereas the measured Pauli limit violation has only minor variations with the displacement field. For $N\geq 3$, the distribution of the displacement field among the layers could be intertwined with the correlation effects in the system, and a spontaneous internal displacement field could in principle exist even when no external displacement field is applied. Slight discrepancies among the twist angles between the layers could introduce further corrections to the displacement field effects. Numerically accounting for these aspects in our calculation is beyond the scope of this work and we hope these issues will be clarified by future research.

The discovery of superconductivity in all members of the magic-angle family from $N=2$ to $N=5$ has profound implications on its mechanism. The presence of superconductivity regardless of the number of layers implies that the peculiar flat band that all these systems share likely plays a crucial role in forming robust superconductivity. Moreover, the $C_{2z}T$ symmetry that all these structures share could also be an important ingredient behind the robustness of the superconductivity, as most other moir\'e systems studied so far lacking such symmetry do not seem to be robust superconductors. These findings put strong constraints on the theories for the possible underlying mechanisms for the unusually strong superconductivity in this moiré family.

\section*{Acknowledgements}
We thank Ashvin Vishwanath, Eslam Khalaf, and Patrick Ledwidth for fruitful discussions. 

This work has been primarily supported by the US Department of Energy (DOE), Office of Basic Energy Sciences (BES), Division of Materials Sciences and Engineering under Award DE-SC0001819 (J.M.P. and S.S.). Help with transport measurements and data analysis were supported by the National Science Foundation (DMR-1809802), and the STC Center for Integrated Quantum Materials (NSF Grant No. DMR-1231319) (Y.C.). Help with device fabrication was supported by the Air Force Office of Scientific Research (AFOSR) 2DMAGIC MURI FA9550-19-1-0390 (L.X.). P.J-H acknowledges support from the Gordon and Betty Moore Foundation's EPiQS Initiative through Grant GBMF9643. P.J-H acknowledges partial support by the Fundación Ramon Areces and the CIFAR Quantum Materials program. The development of new nanofabrication and characterization techniques enabling this work has been supported by the US DOE Office of Science, BES, under award DE‐SC0019300. K.W. and T.T. acknowledge support from the Elemental Strategy Initiative conducted by the MEXT, Japan, Grant Number JPMXP0112101001, JSPS KAKENHI Grant Numbers JP20H00354 and the CREST(JPMJCR15F3), JST. This work made use of the Materials Research Science and Engineering Center Shared Experimental Facilities supported by the National Science Foundation (DMR-0819762) and of Harvard's Center for Nanoscale Systems, supported by the NSF (ECS-0335765). 


\begin{thebibliography}{10}
\expandafter\ifx\csname url\endcsname\relax
  \def\url#1{\texttt{#1}}\fi
\expandafter\ifx\csname urlprefix\endcsname\relax\def\urlprefix{URL }\fi
\providecommand{\bibinfo}[2]{#2}
\providecommand{\eprint}[2][]{\url{#2}}

\bibitem{cao_correlated_2018}
\bibinfo{author}{Cao, Y.} \emph{et~al.}
\newblock \bibinfo{title}{Correlated insulator behaviour at half-filling in
  magic-angle graphene superlattices}.
\newblock \emph{\bibinfo{journal}{Nature}} \textbf{\bibinfo{volume}{556}},
  \bibinfo{pages}{80--84} (\bibinfo{year}{2018}).

\bibitem{cao_unconventional_2018}
\bibinfo{author}{Cao, Y.} \emph{et~al.}
\newblock \bibinfo{title}{Unconventional superconductivity in magic-angle
  graphene superlattices}.
\newblock \emph{\bibinfo{journal}{Nature}} \textbf{\bibinfo{volume}{556}},
  \bibinfo{pages}{43--50} (\bibinfo{year}{2018}).

\bibitem{andrei_marvels_2021}
\bibinfo{author}{Andrei, E.~Y.} \emph{et~al.}
\newblock \bibinfo{title}{The marvels of moiré materials}.
\newblock \emph{\bibinfo{journal}{Nature Reviews Materials}}
  \textbf{\bibinfo{volume}{6}}, \bibinfo{pages}{201--206}
  (\bibinfo{year}{2021}).

\bibitem{park_tunable_2021}
\bibinfo{author}{Park, J.~M.}, \bibinfo{author}{Cao, Y.},
  \bibinfo{author}{Watanabe, K.}, \bibinfo{author}{Taniguchi, T.} \&
  \bibinfo{author}{Jarillo-Herrero, P.}
\newblock \bibinfo{title}{Tunable strongly coupled superconductivity in
  magic-angle twisted trilayer graphene}.
\newblock \emph{\bibinfo{journal}{Nature}} \textbf{\bibinfo{volume}{590}},
  \bibinfo{pages}{249--255} (\bibinfo{year}{2021}).

\bibitem{hao_electric_2021}
\bibinfo{author}{Hao, Z.} \emph{et~al.}
\newblock \bibinfo{title}{Electric field–tunable superconductivity in
  alternating-twist magic-angle trilayer graphene}.
\newblock \emph{\bibinfo{journal}{Science}} \textbf{\bibinfo{volume}{371}},
  \bibinfo{pages}{1133--1138} (\bibinfo{year}{2021}).

\bibitem{khalaf_magic_2019}
\bibinfo{author}{Khalaf, E.}, \bibinfo{author}{Kruchkov, A.~J.},
  \bibinfo{author}{Tarnopolsky, G.} \& \bibinfo{author}{Vishwanath, A.}
\newblock \bibinfo{title}{Magic angle hierarchy in twisted graphene
  multilayers}.
\newblock \emph{\bibinfo{journal}{Physical Review B}}
  \textbf{\bibinfo{volume}{100}}, \bibinfo{pages}{085109}
  (\bibinfo{year}{2019}).

\bibitem{chen_evidence_2019}
\bibinfo{author}{Chen, G.} \emph{et~al.}
\newblock \bibinfo{title}{Evidence of a gate-tunable {Mott} insulator in a
  trilayer graphene moiré superlattice}.
\newblock \emph{\bibinfo{journal}{Nature Physics}}
  \textbf{\bibinfo{volume}{15}}, \bibinfo{pages}{237} (\bibinfo{year}{2019}).

\bibitem{burg_correlated_2019}
\bibinfo{author}{Burg, G.~W.} \emph{et~al.}
\newblock \bibinfo{title}{Correlated {Insulating} {States} in {Twisted}
  {Double} {Bilayer} {Graphene}}.
\newblock \emph{\bibinfo{journal}{Physical Review Letters}}
  \textbf{\bibinfo{volume}{123}}, \bibinfo{pages}{197702}
  (\bibinfo{year}{2019}).

\bibitem{shen_correlated_2020}
\bibinfo{author}{Shen, C.} \emph{et~al.}
\newblock \bibinfo{title}{Correlated states in twisted double bilayer
  graphene}.
\newblock \emph{\bibinfo{journal}{Nature Physics}}
  \textbf{\bibinfo{volume}{16}}, \bibinfo{pages}{520--525}
  (\bibinfo{year}{2020}).

\bibitem{cao_tunable_2020}
\bibinfo{author}{Cao, Y.} \emph{et~al.}
\newblock \bibinfo{title}{Tunable correlated states and spin-polarized phases
  in twisted bilayer–bilayer graphene}.
\newblock \emph{\bibinfo{journal}{Nature}} \textbf{\bibinfo{volume}{583}},
  \bibinfo{pages}{215--220} (\bibinfo{year}{2020}).

\bibitem{liu_tunable_2020}
\bibinfo{author}{Liu, X.} \emph{et~al.}
\newblock \bibinfo{title}{Tunable spin-polarized correlated states in twisted
  double bilayer graphene}.
\newblock \emph{\bibinfo{journal}{Nature}} \textbf{\bibinfo{volume}{583}},
  \bibinfo{pages}{221--225} (\bibinfo{year}{2020}).

\bibitem{he_symmetry_2020}
\bibinfo{author}{He, M.} \emph{et~al.}
\newblock \bibinfo{title}{Symmetry breaking in twisted double bilayer
  graphene}.
\newblock \emph{\bibinfo{journal}{Nature Physics}} \bibinfo{pages}{1--5}
  (\bibinfo{year}{2020}).

\bibitem{polshyn_electrical_2020}
\bibinfo{author}{Polshyn, H.} \emph{et~al.}
\newblock \bibinfo{title}{Electrical switching of magnetic order in an orbital
  {Chern} insulator}.
\newblock \emph{\bibinfo{journal}{Nature}} \textbf{\bibinfo{volume}{588}},
  \bibinfo{pages}{66--70} (\bibinfo{year}{2020}).

\bibitem{xu_tunable_2021}
\bibinfo{author}{Xu, S.} \emph{et~al.}
\newblock \bibinfo{title}{Tunable van {Hove} singularities and correlated
  states in twisted monolayer–bilayer graphene}.
\newblock \emph{\bibinfo{journal}{Nature Physics}}
  \textbf{\bibinfo{volume}{17}}, \bibinfo{pages}{619--626}
  (\bibinfo{year}{2021}).

\bibitem{chen_electrically_2020}
\bibinfo{author}{Chen, S.} \emph{et~al.}
\newblock \bibinfo{title}{Electrically tunable correlated and topological
  states in twisted monolayer–bilayer graphene}.
\newblock \emph{\bibinfo{journal}{Nature Physics}} \bibinfo{pages}{1--7}
  (\bibinfo{year}{2020}).

\bibitem{regan_mott_2020}
\bibinfo{author}{Regan, E.~C.} \emph{et~al.}
\newblock \bibinfo{title}{Mott and generalized {Wigner} crystal states in {WSe}
  2 /{WS} 2 moiré superlattices}.
\newblock \emph{\bibinfo{journal}{Nature}} \textbf{\bibinfo{volume}{579}},
  \bibinfo{pages}{359--363} (\bibinfo{year}{2020}).

\bibitem{tang_simulation_2020}
\bibinfo{author}{Tang, Y.} \emph{et~al.}
\newblock \bibinfo{title}{Simulation of {Hubbard} model physics in {WSe} 2
  /{WS} 2 moiré superlattices}.
\newblock \emph{\bibinfo{journal}{Nature}} \textbf{\bibinfo{volume}{579}},
  \bibinfo{pages}{353--358} (\bibinfo{year}{2020}).

\bibitem{wang_correlated_2020}
\bibinfo{author}{Wang, L.} \emph{et~al.}
\newblock \bibinfo{title}{Correlated electronic phases in twisted bilayer
  transition metal dichalcogenides}.
\newblock \emph{\bibinfo{journal}{Nature Materials}}
  \textbf{\bibinfo{volume}{19}}, \bibinfo{pages}{861--866}
  (\bibinfo{year}{2020}).

\bibitem{xu_correlated_2020}
\bibinfo{author}{Xu, Y.} \emph{et~al.}
\newblock \bibinfo{title}{Correlated insulating states at fractional fillings
  of moiré superlattices}.
\newblock \emph{\bibinfo{journal}{Nature}} \textbf{\bibinfo{volume}{587}},
  \bibinfo{pages}{214--218} (\bibinfo{year}{2020}).

\bibitem{jin_stripe_2021}
\bibinfo{author}{Jin, C.} \emph{et~al.}
\newblock \bibinfo{title}{Stripe phases in {WSe2}/{WS2} moiré superlattices}.
\newblock \emph{\bibinfo{journal}{Nature Materials}}
  \textbf{\bibinfo{volume}{20}}, \bibinfo{pages}{940--944}
  (\bibinfo{year}{2021}).

\bibitem{li_continuous_2021}
\bibinfo{author}{Li, T.} \emph{et~al.}
\newblock \bibinfo{title}{Continuous {Mott} transition in semiconductor moiré
  superlattices}.
\newblock \emph{\bibinfo{journal}{Nature}} \textbf{\bibinfo{volume}{597}},
  \bibinfo{pages}{350--354} (\bibinfo{year}{2021}).

\bibitem{li_quantum_2021}
\bibinfo{author}{Li, T.} \emph{et~al.}
\newblock \bibinfo{title}{Quantum anomalous {Hall} effect from intertwined
  moir{\textbackslash}'e bands}.
\newblock \emph{\bibinfo{journal}{arXiv:2107.01796 [cond-mat]}}
  (\bibinfo{year}{2021}).

\bibitem{yankowitz_tuning_2019}
\bibinfo{author}{Yankowitz, M.} \emph{et~al.}
\newblock \bibinfo{title}{Tuning superconductivity in twisted bilayer
  graphene}.
\newblock \emph{\bibinfo{journal}{Science}} \textbf{\bibinfo{volume}{363}},
  \bibinfo{pages}{1059--1064} (\bibinfo{year}{2019}).

\bibitem{lu_superconductors_2019}
\bibinfo{author}{Lu, X.} \emph{et~al.}
\newblock \bibinfo{title}{Superconductors, orbital magnets and correlated
  states in magic-angle bilayer graphene}.
\newblock \emph{\bibinfo{journal}{Nature}} \textbf{\bibinfo{volume}{574}},
  \bibinfo{pages}{653--657} (\bibinfo{year}{2019}).

\bibitem{chen_signatures_2019}
\bibinfo{author}{Chen, G.} \emph{et~al.}
\newblock \bibinfo{title}{Signatures of tunable superconductivity in a trilayer
  graphene moiré superlattice}.
\newblock \emph{\bibinfo{journal}{Nature}} \textbf{\bibinfo{volume}{572}},
  \bibinfo{pages}{215--219} (\bibinfo{year}{2019}).

\bibitem{zhang_correlated_2021}
\bibinfo{author}{Zhang, X.} \emph{et~al.}
\newblock \bibinfo{title}{Correlated {Insulating} {States} and {Transport}
  {Signature} of {Superconductivity} in {Twisted} {Trilayer} {Graphene}
  {Superlattices}}.
\newblock \emph{\bibinfo{journal}{Physical Review Letters}}
  \textbf{\bibinfo{volume}{127}}, \bibinfo{pages}{166802}
  (\bibinfo{year}{2021}).

\bibitem{cao_pauli-limit_2021}
\bibinfo{author}{Cao, Y.}, \bibinfo{author}{Park, J.~M.},
  \bibinfo{author}{Watanabe, K.}, \bibinfo{author}{Taniguchi, T.} \&
  \bibinfo{author}{Jarillo-Herrero, P.}
\newblock \bibinfo{title}{Pauli-limit violation and re-entrant
  superconductivity in moiré graphene}.
\newblock \emph{\bibinfo{journal}{Nature}} \textbf{\bibinfo{volume}{595}},
  \bibinfo{pages}{526--531} (\bibinfo{year}{2021}).

\bibitem{suarez_morell_flat_2010}
\bibinfo{author}{Suárez~Morell, E.}, \bibinfo{author}{Correa, J.~D.},
  \bibinfo{author}{Vargas, P.}, \bibinfo{author}{Pacheco, M.} \&
  \bibinfo{author}{Barticevic, Z.}
\newblock \bibinfo{title}{Flat bands in slightly twisted bilayer graphene:
  {Tight}-binding calculations}.
\newblock \emph{\bibinfo{journal}{Physical Review B}}
  \textbf{\bibinfo{volume}{82}}, \bibinfo{pages}{121407}
  (\bibinfo{year}{2010}).

\bibitem{bistritzer_moire_2011}
\bibinfo{author}{Bistritzer, R.} \& \bibinfo{author}{MacDonald, A.~H.}
\newblock \bibinfo{title}{Moiré bands in twisted double-layer graphene}.
\newblock \emph{\bibinfo{journal}{Proceedings of the National Academy of
  Sciences}} \textbf{\bibinfo{volume}{108}}, \bibinfo{pages}{12233--12237}
  (\bibinfo{year}{2011}).

\bibitem{lopes_dos_santos_continuum_2012}
\bibinfo{author}{Lopes~dos Santos, J. M.~B.}, \bibinfo{author}{Peres, N. M.~R.}
  \& \bibinfo{author}{Castro~Neto, A.~H.}
\newblock \bibinfo{title}{Continuum model of the twisted graphene bilayer}.
\newblock \emph{\bibinfo{journal}{Physical Review B}}
  \textbf{\bibinfo{volume}{86}}, \bibinfo{pages}{155449}
  (\bibinfo{year}{2012}).

\bibitem{xie_spectroscopic_2019}
\bibinfo{author}{Xie, Y.} \emph{et~al.}
\newblock \bibinfo{title}{Spectroscopic signatures of many-body correlations in
  magic-angle twisted bilayer graphene}.
\newblock \emph{\bibinfo{journal}{Nature}} \textbf{\bibinfo{volume}{572}},
  \bibinfo{pages}{101--105} (\bibinfo{year}{2019}).

\bibitem{wong_cascade_2020}
\bibinfo{author}{Wong, D.} \emph{et~al.}
\newblock \bibinfo{title}{Cascade of electronic transitions in magic-angle
  twisted bilayer graphene}.
\newblock \emph{\bibinfo{journal}{Nature}} \textbf{\bibinfo{volume}{582}},
  \bibinfo{pages}{198--202} (\bibinfo{year}{2020}).

\bibitem{zondiner_cascade_2020}
\bibinfo{author}{Zondiner, U.} \emph{et~al.}
\newblock \bibinfo{title}{Cascade of phase transitions and {Dirac} revivals in
  magic-angle graphene}.
\newblock \emph{\bibinfo{journal}{Nature}} \textbf{\bibinfo{volume}{582}},
  \bibinfo{pages}{203--208} (\bibinfo{year}{2020}).

\bibitem{park_flavour_2021}
\bibinfo{author}{Park, J.~M.}, \bibinfo{author}{Cao, Y.},
  \bibinfo{author}{Watanabe, K.}, \bibinfo{author}{Taniguchi, T.} \&
  \bibinfo{author}{Jarillo-Herrero, P.}
\newblock \bibinfo{title}{Flavour {Hund}’s coupling, {Chern} gaps and charge
  diffusivity in moiré graphene}.
\newblock \emph{\bibinfo{journal}{Nature}} \textbf{\bibinfo{volume}{592}},
  \bibinfo{pages}{43--48} (\bibinfo{year}{2021}).

\bibitem{oh_evidence_2021}
\bibinfo{author}{Oh, M.} \emph{et~al.}
\newblock \bibinfo{title}{Evidence for unconventional superconductivity in
  twisted bilayer graphene}.
\newblock \emph{\bibinfo{journal}{Nature}} \textbf{\bibinfo{volume}{600}},
  \bibinfo{pages}{240--245} (\bibinfo{year}{2021}).

\bibitem{pierce_unconventional_2021}
\bibinfo{author}{Pierce, A.~T.} \emph{et~al.}
\newblock \bibinfo{title}{Unconventional sequence of correlated {Chern}
  insulators in magic-angle twisted bilayer graphene}.
\newblock \emph{\bibinfo{journal}{Nature Physics}}
  \textbf{\bibinfo{volume}{17}}, \bibinfo{pages}{1210--1215}
  (\bibinfo{year}{2021}).

\bibitem{xie_fractional_2021}
\bibinfo{author}{Xie, Y.} \emph{et~al.}
\newblock \bibinfo{title}{Fractional {Chern} insulators in magic-angle twisted
  bilayer graphene}.
\newblock \emph{\bibinfo{journal}{Nature}} \textbf{\bibinfo{volume}{600}},
  \bibinfo{pages}{439--443} (\bibinfo{year}{2021}).

\bibitem{rodan-legrain_highly_2021}
\bibinfo{author}{Rodan-Legrain, D.} \emph{et~al.}
\newblock \bibinfo{title}{Highly tunable junctions and non-local {Josephson}
  effect in magic-angle graphene tunnelling devices}.
\newblock \emph{\bibinfo{journal}{Nature Nanotechnology}} \bibinfo{pages}{1--7}
  (\bibinfo{year}{2021}).

\bibitem{de_vries_gate-defined_2021}
\bibinfo{author}{de~Vries, F.~K.} \emph{et~al.}
\newblock \bibinfo{title}{Gate-defined {Josephson} junctions in magic-angle
  twisted bilayer graphene}.
\newblock \emph{\bibinfo{journal}{Nature Nanotechnology}}
  \textbf{\bibinfo{volume}{16}}, \bibinfo{pages}{760--763}
  (\bibinfo{year}{2021}).

\bibitem{cao_strange_2020}
\bibinfo{author}{Cao, Y.} \emph{et~al.}
\newblock \bibinfo{title}{Strange {Metal} in {Magic}-{Angle} {Graphene} with
  near {Planckian} {Dissipation}}.
\newblock \emph{\bibinfo{journal}{Physical Review Letters}}
  \textbf{\bibinfo{volume}{124}}, \bibinfo{pages}{076801}
  (\bibinfo{year}{2020}).

\bibitem{polshyn_large_2019}
\bibinfo{author}{Polshyn, H.} \emph{et~al.}
\newblock \bibinfo{title}{Large linear-in-temperature resistivity in twisted
  bilayer graphene}.
\newblock \emph{\bibinfo{journal}{Nature Physics}}
  \textbf{\bibinfo{volume}{15}}, \bibinfo{pages}{1011--1016}
  (\bibinfo{year}{2019}).

\bibitem{dean_boron_2010}
\bibinfo{author}{Dean, C.~R.} \emph{et~al.}
\newblock \bibinfo{title}{Boron nitride substrates for high-quality graphene
  electronics}.
\newblock \emph{\bibinfo{journal}{Nature Nanotechnology}}
  \textbf{\bibinfo{volume}{5}}, \bibinfo{pages}{722--726}
  (\bibinfo{year}{2010}).

\bibitem{qin_-plane_2021}
\bibinfo{author}{Qin, W.} \& \bibinfo{author}{MacDonald, A.~H.}
\newblock \bibinfo{title}{In-{Plane} {Critical} {Magnetic} {Fields} in
  {Magic}-{Angle} {Twisted} {Trilayer} {Graphene}}.
\newblock \emph{\bibinfo{journal}{Physical Review Letters}}
  \textbf{\bibinfo{volume}{127}}, \bibinfo{pages}{097001}
  (\bibinfo{year}{2021}).

\bibitem{cao_nematicity_2021}
\bibinfo{author}{Cao, Y.} \emph{et~al.}
\newblock \bibinfo{title}{Nematicity and competing orders in superconducting
  magic-angle graphene}.
\newblock \emph{\bibinfo{journal}{Science}} \textbf{\bibinfo{volume}{372}},
  \bibinfo{pages}{264--271} (\bibinfo{year}{2021}).

\bibitem{ledwith_tb_2021}
\bibinfo{author}{Ledwith, P.~J.} \emph{et~al.}
\newblock \bibinfo{title}{{TB} or not {TB}? {Contrasting} properties of twisted
  bilayer graphene and the alternating twist \$n\$-layer structures (\$n=3, 4,
  5, {\textbackslash}dots\$)}.
\newblock \emph{\bibinfo{journal}{arXiv:2111.11060 [cond-mat]}}
  (\bibinfo{year}{2021}).

\bibitem{lake_reentrant_2021}
\bibinfo{author}{Lake, E.} \& \bibinfo{author}{Senthil, T.}
\newblock \bibinfo{title}{Reentrant superconductivity through a quantum
  {Lifshitz} transition in twisted trilayer graphene}.
\newblock \emph{\bibinfo{journal}{Physical Review B}}
  \textbf{\bibinfo{volume}{104}}, \bibinfo{pages}{174505}
  (\bibinfo{year}{2021}).

\end{thebibliography}
\end{document}